\documentclass[
 reprint,
 amsmath,amssymb,
 aps,
 prl,
]{revtex4-1}

\usepackage{graphicx}
\usepackage{bm}
\usepackage{chemfig}
\newcommand{\Gg}{\mathcal{G}}
\usepackage{nicefrac}
\usepackage{subfig}
\usepackage{enumitem}

\begin{document}

\preprint{APS/123-QED}

\title{Three-dimensional organic Dirac-line materials due to nonsymmorphic symmetry: a  data mining approach}

\author{R. Matthias Geilhufe$^1$}
\email{Geilhufe@kth.se}
\affiliation{$^1$Nordita, Center for Quantum Materials, KTH Royal Institute of Technology and Stockholm University, Roslagstullsbacken 23, SE-106 91 Stockholm, Sweden\\
$^2$Department of Physics and Astronomy, Uppsala University, Box 516, S-751 20 Uppsala, Sweden}
\author{Adrien Bouhon$^2$}
\affiliation{$^1$Nordita, Center for Quantum Materials, KTH Royal Institute of Technology and Stockholm University, Roslagstullsbacken 23, SE-106 91 Stockholm, Sweden\\
$^2$Department of Physics and Astronomy, Uppsala University, Box 516, S-751 20 Uppsala, Sweden}
\author{Stanislav S. Borysov$^1$}
\affiliation{$^1$Nordita, Center for Quantum Materials, KTH Royal Institute of Technology and Stockholm University, Roslagstullsbacken 23, SE-106 91 Stockholm, Sweden\\
$^2$Department of Physics and Astronomy, Uppsala University, Box 516, S-751 20 Uppsala, Sweden}
\author{Alexander V. Balatsky$^1$}
\affiliation{$^1$Nordita, Center for Quantum Materials, KTH Royal Institute of Technology and Stockholm University, Roslagstullsbacken 23, SE-106 91 Stockholm, Sweden\\
$^2$Department of Physics and Astronomy, Uppsala University, Box 516, S-751 20 Uppsala, Sweden}

\date{\today}

\begin{abstract}
A data mining study of electronic Kohn-Sham band structures was performed to identify Dirac materials within the Organic Materials Database (OMDB). Out of that, the three-dimensional organic crystal 5,6-bis(trifluoromethyl)-2-methoxy-1$H$-1,3-diazepine was found to host different Dirac line-nodes within the band structure. From a group theoretical analysis, it is possible to distinguish between Dirac line-nodes occurring due to twofold degenerate energy levels protected by the monoclinic crystalline symmetry and twofold degenerate accidental crossings protected by the topology of the electronic band structure. The obtained results can be generalized to all materials having the space group $P2_1/c$ (No. 14, $C^5_{2h}$) by introducing three distinct topological classes. 
\end{abstract}

\maketitle

\section{\label{sec:Intro}Introduction}
The investigation of Dirac materials, i.e., materials where the low-energy excitations behave as massless Dirac fermions, became of increasing interest during the past decades~\cite{wehling2014}. Among the most prominent examples are the two-dimensional material graphene~\cite{abergel2010properties}, bulk topological insulators~\cite{fu2007topological} such as Pb$ _x$Sn$_{1-x}$Te~\cite{tanaka2012experimental,geilhufe2015effect,hsieh2012topological} or Bi$_2$Se$_3$~\cite{henk2012complex,rauch2014dual,chen2009experimental} hosting massless Dirac fermions on the surface and Weyl semimetals such as TaAs~\cite{xu2015discovery,lv2015experimental}. 

Recently, so-called Dirac-line materials where identified, for example, in the inversion-symmetric crystal Cu$_3$N~\cite{Kim2015}, the antiperovskite Cu$_3$PdN~\cite{Yu2015}, the noncentrosymmetric pnictides CaAg$X$ ($X$ = P, As)~\cite{Yamakage2016} and in three-dimensional graphene networks~\cite{Weng2015}. Within these materials, a linear crossing of energy bands is protected along a path within the Brillouin zone due to the crystalline symmetry. By introducing so-called type-II non-symmorphic symmetry elements, which are characterized by a partial translation orthogonal to the invariant space of the point group operation, Yang \textit{et al.}~\cite{yang2016} were able to show that point nodes can be expected if type-II non-symmorphic rotation symmetries are present, whereas line nodes are expected with type-II non-symmorphic mirror symmetries. The protection of nodal lines by glide mirror symmetries for systems including spin-orbit coupling and one-fold degenerate bands was investigated by Bzdu\v{s}ek \textit{et al.} \cite{bzdusek2016}. A study of point- and line-nodes in layer groups was reported by Wieder and Kane~\cite{wieder2016spin}. Taking into account the whole symmetry group of the crystal, the degeneracy of an energy level is equal to the dimension of the associated irreducible representation, which is a direct consequence of the Wigner-Eckart-theorem~\cite{cornwell1984group,dresselhaus2007group}. Various possibilities of observing Dirac-like or even more exotic crossings due to higher dimensional irreducible representations occuring at high-symmetry points within the Brillouin zone were discussed, e.g., by Bradlyn \textit{et al.}~\cite{bradlyn2016beyond} and Wieder \textit{et al.}~\cite{PhysRevLett.116.186402}. 

In comparison to inorganic materials, organic Dirac materials are investigated rarely. However, these materials show an interesting perspective for technological applications, especially, for electronic devices~\cite{forrest2007introduction,klauk2006organic,kelley2004recent,shaw2001organic}. 
To identify Dirac materials in the class of three-dimensional (3D) organic crystals, a data mining study was performed on the basis of more than 5000 electronic Kohn-Sham band structures stored within the Organic Materials Database (OMDB)~\cite{geilhufe2016a}. Techniques of data mining represent a modern approach within materials science sometimes referred to as materials informatics~\cite{rodgers2006materials,ferris2007materials}. Recently, data mining was successfully applied, for example, for the search of stable nitride perovskites~\cite{sarmiento2015prediction} or bulk topological insulators~\cite{klintenberg2014computational}. 

In the following, we discuss the three-dimensional organic crystal 5,6-bis(trifluoromethyl)-2-methoxy-1$H$-1,3-diazepine which represents a 3D organic Dirac-line material where the Dirac line occurs close enough to the Fermi level to be tuned by $p$ doping, by producing layered structures or within a field-effect setup. Although details about the synthesis of the material are given in Ref.~\cite{reisinger20041}, an investigation of the electronic structure has not been reported yet. From a group theoretical investigation of the monoclinic space group, we distinguish between two types of Dirac lines found within the electronic structure. The first type results from two-fold degenerate levels protected by the crystalline symmetry. The second class belongs to topologically protected accidental crossings which have to occur along certain paths within the Brillouin zone due to band characteristics and their compatibility relations. We discuss the origin of those crossing and generalize the statement to all materials having the space group $P2_1/c$ (or equivalent).

\section{Data mining\label{sec4}}
The data mining was performed on more than 5000 calculated Kohn-Sham band structures for organic crystals stored within the OMDB, accessible at \url{http://omdb.diracmaterials.org}~\cite{geilhufe2016a}. As pointed out in Ref.~\cite{geilhufe2016a}, the band gap distribution of three-dimensional organic crystals can be represented by a Gaussian with a mean of $2.9$~eV, i.e., most organic crystals are large band gap insulators. However, doping of organic materials is widely studied opening the opportunity of shifting the Fermi level into the valence or conduction band~\cite{lussem2013doping}. Therefore, we searched for linear crossings in a neighborhood of $0.1$~eV from the lowest unoccupied electronic state (LUES) and the highest occupied electronic state (HOES). As a ``hard'' search criteria, we were looking for tiny energy gaps of the size of less than $0.01$~eV: first, to search for isolated Dirac crossings and, second, to introduce a numerical tolerance since the band structure calculations were performed along a discrete mesh and the crossing point might not be represented. 

In a second step, we applied a pattern matching algorithm based on the moving window approach to range selected band structures according to their similarity to a linear crossing pattern. The algorithm scans through each pair of adjacent bands of the materials selected in the first step by using a moving window of characteristic pattern size (0.4 reciprocal length units in the current paper, the number was chosen empirically). Then, the energy values within each window are scaled linearly to have the same maximum and minimum values as the pattern. Finally, the average Euclidean distance (i.e., the root mean square error) between the re-scaled energy values and the pattern values with the same momentum is calculated and used as a similarity measure. Although the algorithm represents the most basic approach to pattern matching, leading to a relatively high number of false positive results, it helped to reduce overall data analysis time by picking up the most promising materials first.
Following this procedure, the material 5,6-bis(trifluoromethyl)-2-methoxy-1$H$-1,3-diazepine was found to show an isolated linear crossing at a distance of $-70$~meV from the HOES along the path $\overline{\Gamma Z}$. The full band structure of the material is illustrated in Fig.~\ref{f5a}. The chosen path within the Brillouin zone is shown in Fig.~\ref{f5c}.

\begin{table}[b!]
\begin{tabular}{lcccc}
\hline
\hline
& $a$ & $b$ & $c$  & $\beta$ \\
\hline
experimental~\cite{reisinger20041} & $9.642$ & $10.277$ & $10.081$ & $91.89^\circ$\\
computational & $9.204$ & $9.876$ & $9.724$ & $92.18^\circ$ \\
\hline
\hline
\end{tabular}
\caption{C$_{8}$H$_{6}$F$_6$N$_2$O lattice constants (\AA)\label{T3}}
\end{table}

%
%
\begin{figure}[b]
\subfloat[The C$_{8}$H$_{6}$F$_6$N$_2$O molecule\label{f3a}]{\includegraphics[height=5.2cm]{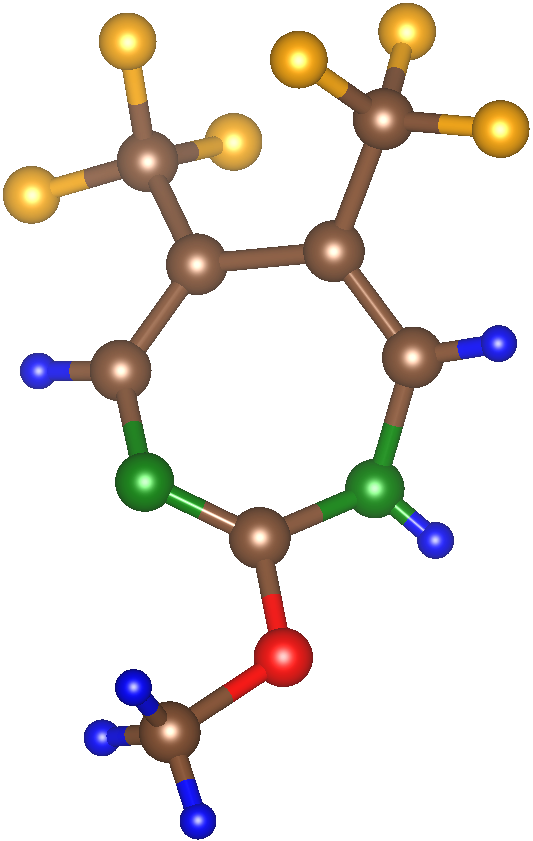}}
\subfloat[The monoclinic unit cell\label{f3b}]{\includegraphics[height=5.2cm]{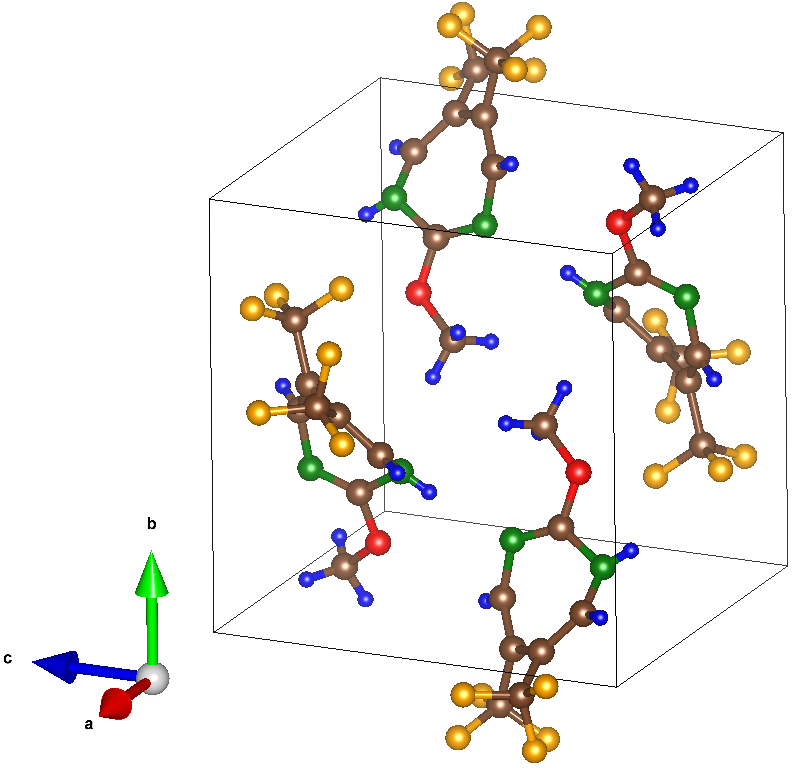}}\\
\caption{Illustration of the monoclinic crystal structure containing four 5,6-Bis(trifluoromethyl)-2-methoxy- 1H-1,3-diazepine molecules in the unit cell~\cite{momma2011vesta}. The colors indicate: brown for carbon, blue for hydrogen, yellow for fluorine, green for nitrogen and red for oxygen.}
\end{figure}
\section{\textit{ab initio} investigation of 5,6-bis(trifluoromethyl)-2-methoxy-1$H$-1,3-diazepine\label{sec5}}
%
%
\begin{figure*}[t!]
\raggedright
\subfloat[Band structure in the whole Brillouin zone.\label{f5a}]{\includegraphics[height=5.1cm]{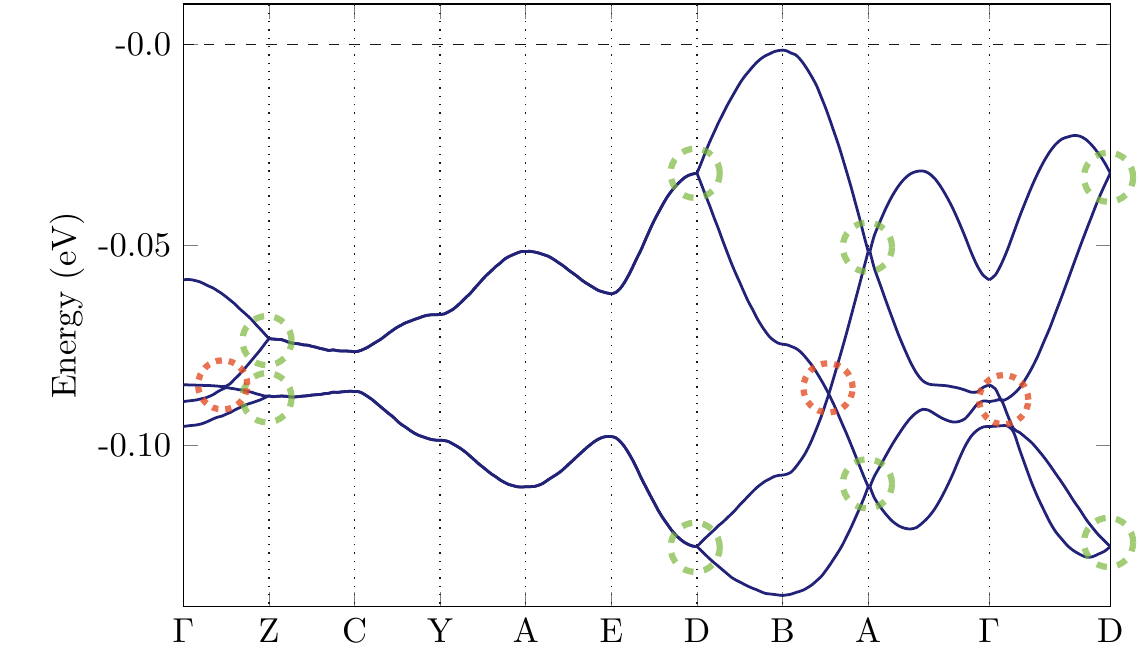}}\hspace{0.5cm}
\subfloat[Energy dispersion along $\overline{\Gamma B}$, $\overline{\Gamma Y}$, $\overline{\Gamma Z}$ and irreducible representations\label{f5b}]{\includegraphics[height=5.1cm]{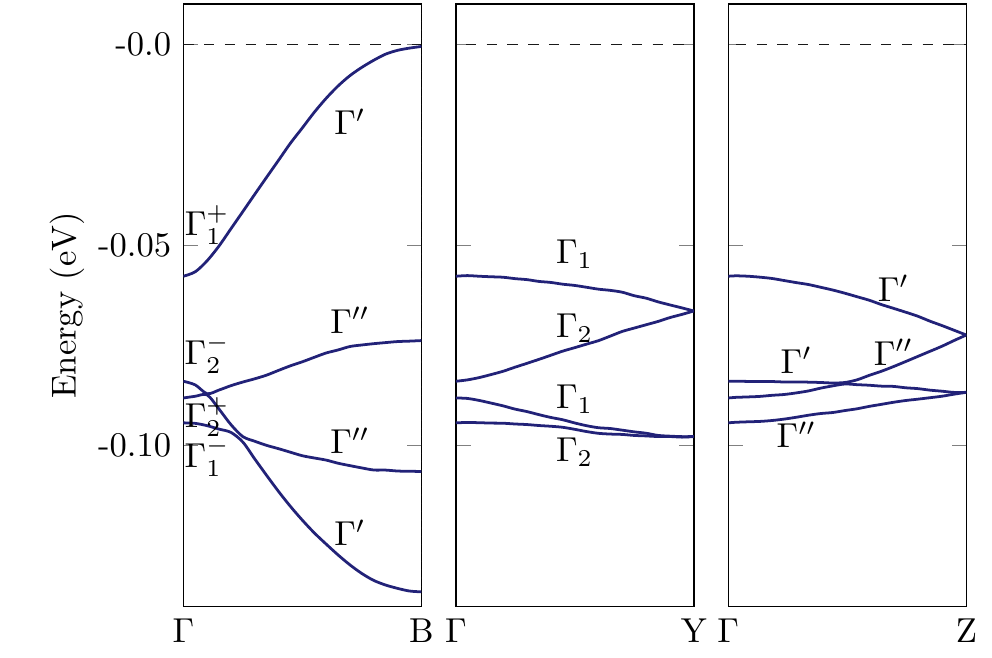}}\\
\subfloat[Brillouin zone path for band structure calculation in Fig.~\ref{f5a}\label{f5c}]
{\includegraphics[height=4.3cm]{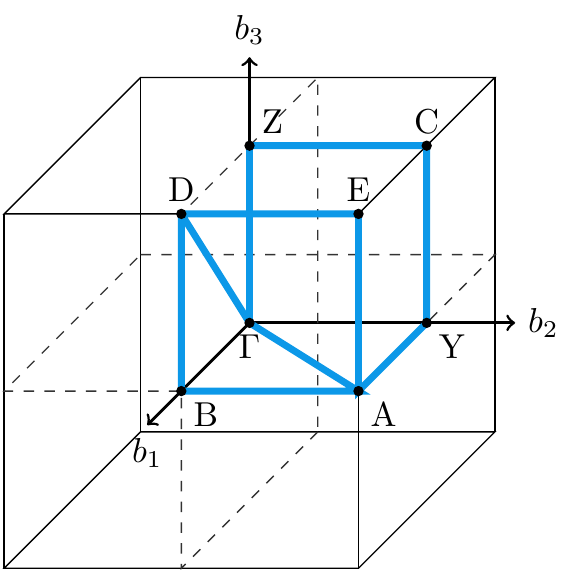}}
\subfloat[twofold degenerate Dirac lines due to crystalline symmetry\label{f5d}]
{\includegraphics[height=4.3cm]{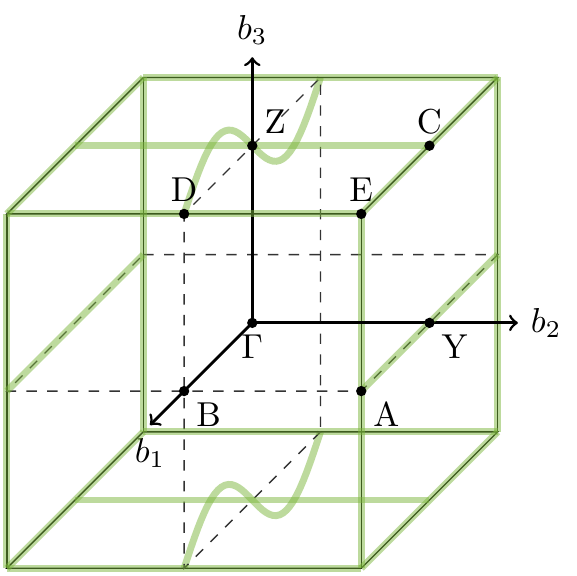}}
\subfloat[Topologically protected Dirac line\label{f5e}]
{\includegraphics[height=4.3cm]{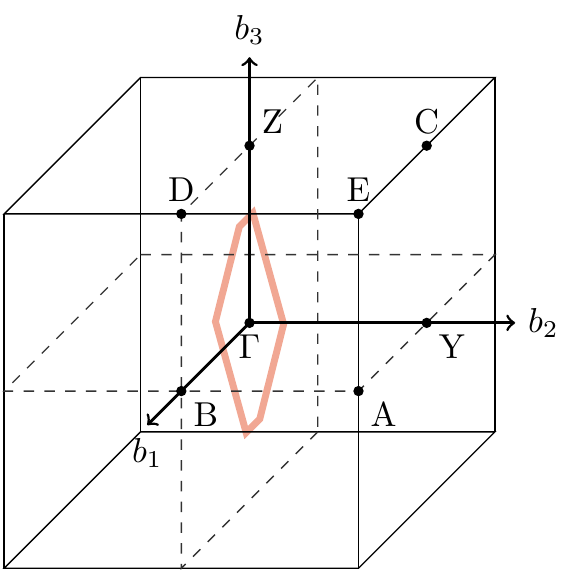}}
\subfloat[3D plot of the energy dispersion around the $D$ point within the $\vec{b}_2$-$\vec{b}_3$-plane\label{f5f}]
{\includegraphics[height=4.3cm]{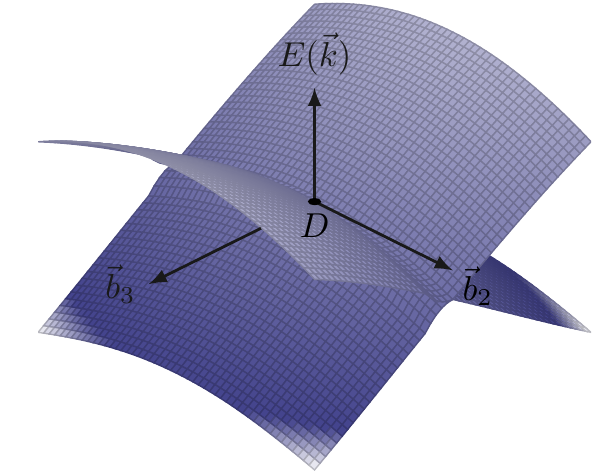}}
\caption{Electronic structure and Brillouin zone (in units of reciprocal lattice vectors) of 5,6-bis(trifluoromethyl)-2-methoxy-1$H$-1,3-diazepine.\label{f5}}
\end{figure*}
5,6-bis(trifluoromethyl)-2-methoxy-1$H$-1,3-diazepine has the chemical sum formula C$_{8}$H$_{6}$F$_6$N$_2$O. The monoclinic unit cell of the crystal contains four C$_{8}$H$_{6}$F$_6$N$_2$O molecules and is shown in Fig.~\ref{f3b}. The lattice vectors are given by
\begin{align}
\vec{a}_1 &= \left(a,0,0\right),\\
\vec{a}_2 &= \left(0,b,0\right),\\
\vec{a}_3 &= \left(c \cos{\left(\beta\right)},0,c \sin{\left(\beta\right)}\right).
\end{align}
The lattice constants $a$, $b$ and $c$ as well as the angle $\beta$ are given in Table \ref{T3}. To obtain a deeper insight into the electronic structure, \textit{ab initio} calculations in the framework of the density functional theory~\cite{Hohenberg1964,Kohn1965,Jones2015} based on the pseudopotential projector augmented-wave method~\cite{hamann1979norm,blochl1994projector,pseudo1,pseudo2,kresse1999ultrasoft} where performed, as implemented in the Vienna \textit{Ab initio} Simulation Package (\textsc{vasp})~\cite{vasp1,vasp2,vasp3} and the \textsc{quantum espresso} code~\cite{qespresso}. While using both codes, the exchange-correlation functional was approximated by the generalized gradient approximation according to Perdew, Burke and Ernzerhof~\cite{perdew1996}. The structural information of the C$_{8}$H$_{6}$F$_6$N$_2$O crystal was taken from the Crystallography Open Database (COD)~\cite{merkys2016cod,gravzulis2015computing,gravzulis2012crystallography,gravzulis2009crystallography,downs2003american} which is available online at \url{http://crystallography.net} and transformed into capable input files for \textsc{vasp}  by applying the Pymatgen package~\cite{Ong2013314}. 

\textsc{vasp} was mainly used for checking the change of the band structure under structural optimization. After the optimization, the unit cell shrinked, as can be verified from Table \ref{T3}. However, in comparison to the original cell, the band structure itself did not change qualitatively. The used pseudopotentials were calculated according to~\cite{kresse1999ultrasoft}. The precision flag was set to ``accurate'' meaning that the energy cut-off is given by the $1.3 E_{\text{cut}}^{\text{max}}$, where $E_{\text{cut}}^{\text{max}}$ denotes the maximum of the specified maxima for the cut-off energies within the POTCAR files (e.g. for carbon this value is given by 400~eV). The calculations were performed spin-polarized but without spin-orbit coupling. For the integration in $\vec{k}$-space, an $8\times8\times8$ $\Gamma$-centred mesh according to Monkhorst and Pack~\cite{monkhorst1976special} was chosen for the self-consistent cycle. During the structural optimization, the ionic positions, the cell shape and the volume were allowed to change. The $\vec{k}$-path for the band structure calculations was chosen according to Fig.~\ref{f5c}. \textsc{quantum espresso} was applied to estimate the associated irreducible representations of the energy levels within the band structure. 
The cut-off energy for the wave function was chosen to be 48~Ry and the cut-off energy for the charge density and the potentials was chosen to be 316~Ry. The calculated band structures using \textsc{vasp} and \textsc{quantum espresso} agree well. 

Within the band structure shown in Fig.~\ref{f5a}, linear crossing of energy bands can be found at the $Z$, $D$ and $A$ points of the Brillouin zone as well as along the paths $\overline{\Gamma Z}$, $\overline{BA}$ and $\overline{\Gamma D}$. The two-fold degeneracy at the high-symmetry points $Z$, $D$ and $A$ is kept along the paths $\overline{ZC}$, $\overline{CY}$, $\overline{YA}$, $\overline{AE}$ and $\overline{ED}$. The crossings and degeneracies are discussed in more detail within the following section.

\section{Group theoretical analysis\label{sec2}}
\subsection{The space group $P2_1/c$}
C$_{8}$H$_{6}$F$_6$N$_2$O crystallizes within a monoclinic crystal structure with the space group $P2_1/c$ or No. 14 and $C^5_{2h}$ in terms of the space group number and the Sch\"onflies notation~\cite{schoenflies1923}, respectively. The space group itself (hereinafter denoted by $\Gg$) is an infinite group having the group of pure translations $\mathcal{T}$ as an infinite and normal subgroup. The point group of the lattice $\Gg_0$ is given by $2/m$ ($C_{2h}$), which is the group of all rotational parts of the space group elements. The factor group $\Gg/\mathcal{T}$ is isomorphic to $2/m$ and the coset representatives are given by
\begin{align}
  T_1 &= (E,0,0,0),\label{eqt1} \\
  T_2 &= (I,0,0,0), \\
  T_3 &= (C_{2y},0,\nicefrac{1}{2},\nicefrac{1}{2}), \label{eqt3}\\
  T_4 &= (IC_{2y},0,-\nicefrac{1}{2},-\nicefrac{1}{2}) \label{eqt4},
\end{align}
where $E$ denotes the identity, $I$ the inversion ($x\rightarrow -x$, $y\rightarrow -y$, $z\rightarrow -z$), $C_{2y}$ a two-fold rotation about the $y$-axis ($x\rightarrow -x$, $z\rightarrow -z$) and $IC_{2y}$ a reflection at the $xy$-plane ($y\rightarrow -y$). Since $2/m$ is an Abelian group, each of the four elements forms a class on its own and four one-dimensional irreducible representations can be found. The character table is shown in Table~\ref{T1}. In the following, the linear crossings found within the band structure and depicted in Fig.~\ref{f5a} will be discussed. Within the figure, two different classes are distinguished, where the crossings highlighted by green dashed circles denote crossings protected by crystalline symmetry and red dotted circles denote crossings protected by the band topology.
\begin{table}[t!]
\subfloat[$2/m$ ($C_{2h}$)\label{T1}]{
\begin{tabular}{lrrrr|cc}
\hline
\hline
& $E$ & $I$ & $C_{2y}$ & $IC_{2y}$  & $\{E, C_{2y} \}$ & $\{E, IC_{2y} \}$ \\
\hline
$\Gamma_1^+$ & 1 & 1 & 1 & 1 & $\Gamma_1$ & $\Gamma'$  \\ 
$\Gamma_1^-$ & 1 & -1 & 1 & -1 & $\Gamma_1$ & $\Gamma''$ \\
$\Gamma_2^+$ & 1 & 1 & -1 & -1 & $\Gamma_2$ & $\Gamma''$ \\
$\Gamma_2^-$ & 1 & -1 &-1 & 1 & $\Gamma_2$ & $\Gamma'$ \\
\hline\hline
\end{tabular}}
\hfill
\subfloat[$m$ ($C_s$), $2$ ($C_2$).\label{T2}]{
\begin{tabular}{lrr}
\hline
\hline
& $E$ & $IC_{2y}$ \\
\hline
$\Gamma'$, $\Gamma_1$ & 1 & 1 \\ 
$\Gamma''$, $\Gamma_2$ & 1 & -1 \\
\hline\hline
\end{tabular}}
\caption{Character tables and compatibility relations of the groups $2/m$ ($C_{2h}$) as well as $m$ ($C_s$) and $2$ ($C_2$), respectively.}
\end{table}

\subsection{Case I: Line nodes protected by crystalline symmetry}
The point group of the lattice $2/m$ is an Abelian group and therefore all irreducible representations of $2/m$ are one-dimensional. In general, each eigenvector of the Hamiltonian belongs to one of the irreducible representations. Since a space group is an infinite group, the number of irreducible representations is also infinite. However, since the group of pure translations is an Abelian normal subgroup, it is possible to denote each irreducible representation by $\Gamma_{\vec{k}}^p$, where $\vec{k}$ is a vector in reciprocal space which at the same time represents an index of irreducible representations of the pure translation group. For each value of $\vec{k}$, different irreducible representations of the space group are possible and therefore a second index $p$ is necessary. At the $\Gamma$-point, i.e., for $\vec{k}=\vec{0}$, the allowed irreducible representations are given by the irreducible representations of $2/m$ itself. Hence, only one-fold degenerate states can be expected. Referring to Fig.~\ref{f5b}, it can be verified for the $\Gamma$-point that each of the irreducible representations is realized among the first four bands below the Fermi level.  

A degeneracy can be expected if irreducible representations with dimension higher than 1 occur or if a pair of irreducible representations is complex conjugate to each other. By applying the program \textsc{repres}~\cite{Aroyo:xo5013} of the Bilbao crystallographic server, available online at \url{http://www.cryst.ehu.es}, we calculated the irreducible representations for the space group at specified $\vec{k}$-points within the Brillouin zone. The study was complemented by constructing a $4\times4$ tight-binding Hamiltonian for four atoms in the unit cell within the 2-center approximation~\cite{Podolskiy2004} as implemented in the \textit{Mathematica} group theory package \textsc{gtpack}~\cite{HergertGeilhufe}. The tight-binding parameters were fitted to the \textit{ab initio} band structures by using least squares. By applying both approaches, it was possible to identify lines of twofold degeneracy as illustrated in Fig.~\ref{f5d}. The result is in perfect agreement with the calculated band structure in Fig.~\ref{f5a}. A plot of the energy dispersion within the $\vec{b}_2$-$\vec{b}_3$-plane in the neighborhood of the $D$-point within the Brillouin zone is shown in Fig.~\ref{f5f}.

\subsection{Case II: Line nodes protected by band topology}
In the second class, line nodes are formed from crossings of bands with different band characteristics, also referred to as accidental crossings~\cite{herring1937}. However, these crossings are forced to occur in the present case. The argument follows from the compatibility relations of the irreducible representations. As discussed before, two-fold degenerate states can be expected to occur at the points $Y=(0,0.5,0)$ and $Z=(0,0,0.5)$ within the Brillouin zone. However, along the path $\overline{\Gamma Y}$, the group of the $\vec{k}$-vector (i.e., the group of all rotational parts $\hat{R}$ with the property $\hat{R}\cdot\vec{k}\simeq \vec{k}$) is given by $\Gg_{\Gamma Y}=\{E,C_{2y}\}$, whereas the group of the $\vec{k}$-vector along the path $\overline{\Gamma Z}$ is given by $\Gg_{\Gamma Z}=\{E,IC_{2y}\}$. Both groups are isomorphic, having the character table shown in Table~\ref{T2}. However, taking a look at the compatibility relations in Table~\ref{T1}, it can be verified that the irreducible representations along the path originate from different irreducible representations at the $\Gamma$-point. Whereas the symmetric (antisymmetric) irreducible representation $\Gamma_1$ ($\Gamma_2$) continues along the path $\overline{\Gamma Y}$ starting from $\Gamma_1^+$ and $\Gamma_1^-$ ($\Gamma_2^+$ and $\Gamma_2^-$) at the $\Gamma$-point, the symmetric (antisymmetric) representation $\Gamma'$ ($\Gamma''$) along the path $\overline{\Gamma Z}$ originates from the representations $\Gamma_1^+$ and $\Gamma_2^-$ ($\Gamma_1^-$ and $\Gamma_2^+$). As soon as the bands approach the $Y$ or the $Z$-point, bands with different band characteristic have to merge pairwise, i.e., $\Gamma_1$ and $\Gamma_2$ at $Y$ and $\Gamma'$ as well as $\Gamma''$ at $Z$. Bands belonging to different irreducible representations along the path $\overline{\Gamma Y}$ or $\overline{\Gamma Z}$ are allowed to cross, whereas bands belonging to equivalent irreducible representations will hybridize. Depending on the ordering of irreducible representations at the $\Gamma$-point, three different topological classes can be defined (we list the $4!/2$ possible permutations which can each be realized on the energy axis in an increasing or decresing order):
 \hspace{-1cm}
\begin{enumerate}[leftmargin=0.3cm] 
\item \textit{Trivial}: Pairs of bands belonging to the irreducible representations $\Gamma_1$ and $\Gamma_2$ as well as $\Gamma'$ and $\Gamma''$ can combine at the Brillouin zone boundary without any reason to cross. This situation can be found for the patterns: $\left(\Gamma_1^+, \Gamma_2^+, \Gamma_1^-, \Gamma_2^-\right)$, $\left(\Gamma_1^+, \Gamma_2^+, \Gamma_2^-, \Gamma_1^-\right)$, $\left(\Gamma_2^+, \Gamma_1^+, \Gamma_1^-, \Gamma_2^-\right)$, $\left(\Gamma_2^+, \Gamma_1^+, \Gamma_2^-, \Gamma_1^-\right)$.
\item \textit{One Dirac line within the mirror plane}: To combine pairs of bands belonging to the irreducible representations $\Gamma'$ and $\Gamma''$ a crossing is forced to take place at the path $\overline{\Gamma Z}$. This crossing is protected by the mirror symmetry which is present within the whole $\vec{b}_1$-$\vec{b}_3$-plane. Thus, a Dirac line can be found. The Dirac line is present for the patterns: 
$\left(\Gamma_1^-,\Gamma_2^+,\Gamma_2^-,\Gamma_1^+\right)$,
$\left(\Gamma_1^-,\Gamma_2^+,\Gamma_1^+,\Gamma_2^-\right)$,
$\left(\Gamma_2^+,\Gamma_1^-,\Gamma_2^-,\Gamma_1^+\right)$,
$\left(\Gamma_2^+,\Gamma_1^-,\Gamma_1^+,\Gamma_2^-\right)$.
\item \textit{One Dirac line crossing the $\overline{\Gamma Y}$-axis and its mirror partner}: In order to combine pairs of bands belonging to the irreducible representations $\Gamma_1$ and $\Gamma_2$  at the $Y$ point, a crossing is forced to occur at the path $\overline{\Gamma Y}$. While only the crossing on the path $\overline{\Gamma Y}$ is protected by $C_{2y}$ rotational symmetry, the whole Dirac line is protected by inversion and time reversal symmetry.  This case can be found for the patterns: $\left(\Gamma_2^-,\Gamma_2^+,\Gamma_1^-,\Gamma_1^+\right)$,
$\left(\Gamma_2^-,\Gamma_2^+,\Gamma_1^+,\Gamma_1^-\right)$,
$\left(\Gamma_2^+,\Gamma_2^-,\Gamma_1^-,\Gamma_1^+\right)$,
$\left(\Gamma_2^+,\Gamma_2^-,\Gamma_1^+,\Gamma_1^-\right)$.
\end{enumerate}
This classification can be generalized by introducing a vector $\vec{n}=(n_1,n_2)$ with the components
\begin{align}
n_1 &= \frac{1}{2}\sum_{i=1}^2 \chi^i_\Gamma(IC_{2y}) \mod 2,\\ n_2&= \frac{1}{2}\sum_{i=1}^2 \chi^i_\Gamma(C_{2y})\mod 2.
\end{align}
The sum runs over the lower two bands within groups of four bands and $\chi^i_\Gamma(C_{2y})$ and $\chi^i_\Gamma(IC_{2y})$ represent the characters of the irreducible representations for the elements $C_{2y}$ and $IC_{2y}$ at the $\Gamma$ point. $\left|\vec{n}\right|=0$ represents the trivial phase and $\left|\vec{n}\right|\neq 0$ the two different line node phases. From the individual values of the $n_i$ it can be verified which line-node phase is present.

In the case of 5,6-bis(trifluoromethyl)-2-methoxy-1$H$-1,3-diazepine a vector of $\vec{n}=(1,0)$ is obtained and a tilted Dirac line can be found along a closed path within the Brillouin zone lying in the $\vec{b}_1$-$\vec{b}_3$-plane as illustrated in Fig.~\ref{f5e}.

\section{conclusion}
We presented the organic crystal 5,6-bis(trifluoromethyl)-2-methoxy-1$H$-1,3-diazepine (C$_{8}$H$_{6}$F$_6$N$_2$O) as a realization of a three-dimensional organic Dirac-line material. The material was found by applying a data mining study within the Organic Materials Database (OMDB). Next to twofold degenerate line nodes along the Brillouin zone boundary protected by the crystalline symmetry, it hosts topologically protected Dirac lines inside the Brillouin zone due to different band characteristics along the 3 directions in reciprocal space. These crossings occur approximately $70$~meV below the highest occupied electronic state. Therefore, a rigid shift of the Fermi level could be possible by alloying, within a layered structure or within a field-effect setup. The presented analysis within the framework of group theory is general for every material with the space group $P2_1/c$ or equivalent, where we distinguished between three topologically different classes: a trivial class and two Dirac-line classes.

\section*{Acknowledgments}
The work is supported by the Swedish Research Council Grant No.~638-2013-9243, the Knut and Alice Wallenberg Foundation, and the European Research Council under the European Union’s Seventh Framework Program (FP/2207-2013)/ERC Grant Agreement No.~DM-321031. Furthermore, the authors acknowledge computational resources from the Max Planck Institute of Microstructure Physics in Halle (Germany) and the Swedish National Infrastructure for Computing (SNIC) at the National Supercomputer Centre at Link\"oping University.

\bibliography{references}
\end{document}